\documentstyle[12pt,axodraw]{article}

\makeatletter
\def\@cite#1#2{\/${}^{\hbox{\scriptsize #1\if@tempswa ,#2\fi})}$}
\makeatother

\begin{document}
\vspace*{3cm}
\pagestyle{empty}

\begin{center}
{\large SIGNATURES OF SUPERSYMMETRIC Q-BALLS}
\end{center}
\vspace*{1cm}

\begin{center}
P.G.Tinyakov \\
{\em Institute for Nuclear Research of the Russian
   Academy of Sciences,}\\
{\em 60th October Anniversary prospect
   7a, Moscow 117312, Russia.}
\end{center}

\vfill
\begin{center} 
ABSTRACT\\
\parbox[t]{15cm}{\baselineskip 14truept In many supersymmetric
extensions of the Standard Model the spectrum of states contains
stable non-topological solitons, Q-balls. If formed in the Early
Universe in sufficient amounts, Q-balls now contribute to cold dark
matter. We discuss their experimental signatures and astrophysical
implications.}
\end{center}
\vfill

\newpage
\paragraph{1. Q-balls in SUSY theories.} In theories where scalar
fields carry a conserved global charge, $Q$, there may exist
non-topological solitons which are stabilized by the charge
conservation\cite{FLS}. Under certain assumptions about the
self-interaction of the scalar fields these solitons, which are called
Q-balls\cite{Coleman}, are absolutely stable. The prototype
model\cite{KSD} containing absolutely stable Q-balls has one complex
scalar field $\phi$ with the potential $V(|\phi|)$ which is
asymptotically flat as shown in Fig.1a. The Q-ball solution has the
form $\phi(t,x) = \exp(i\omega t) \phi(r)$, where $\omega$ and
$\phi(r)$ are found by minimizing the energy in the sector of fixed
charge $Q$. The profile $\phi(r)$ is shown schematically in Fig.1b. In
the interior region $\phi(r)$ satisfies massive free field equation,
$\phi(r) = \phi_0 \sin(\omega r)/\omega r$.  At $r\approx
R=\pi/\omega$ there is a transition region where the field smoothly
goes to zero in a way determined by the shape of the potential at
small fields.  The Q-ball mass $M_Q$ and size $R$, as well as the
parameters $\omega$ and $\phi_0$ are functions of its charge Q. In
particular,
\begin{eqnarray} 
M_{_Q} & = & 4\pi\sqrt{2}/3 M_s Q^{3/4} ,\nonumber \\
R & = & 1/\sqrt{2} M_s^{-1} Q^{1/4},
\label{Q_M,R}
\end{eqnarray}
where $M_s$ is determined the asymptotic value of the potential,
$V\simeq M_s^4$.

The conditions for existence of absolutely stable Q-balls are
naturally satisfied\cite{KSD} in supersymmertic theories with low
energy supersymmetry breaking. The role of conserved charge is played
by the baryon number $B$, while scalar fields carrying the charge are
certain combinations of squark, slepton and Higgs fields associated
with flat directions of scalar potential (for the list of flat
directions of the MSSM see\cite{flat}). The mass parameter $M_s$ is
of the order of the SUSY breaking scale. For definiteness we take
$M_s=1$ TeV. The condition of absolute stability $M_{_Q}<Qm_p$, where
$m_p$ is the proton mass, is satisfied for charges
$Q>10^{15}(M_s/1\mbox{TeV})^4$.

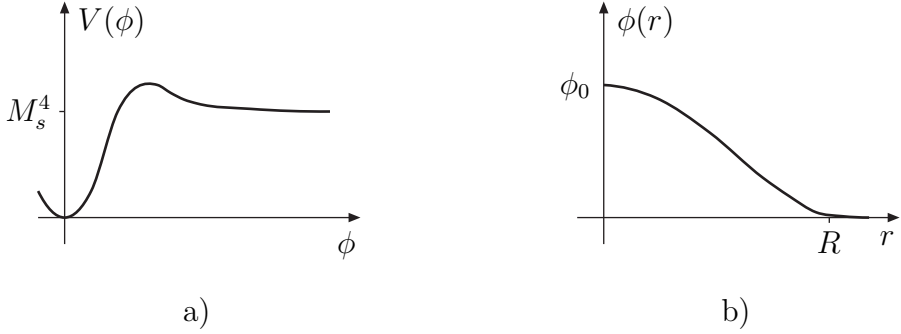
\begin{figure}
\begin{center}
\begin{picture}(200,200)(0,0) 
\SetOffset(20,10)
\LongArrow(0,10)(120,10)
\LongArrow(10,0)(10,90)
\Text(15,90)[tl]{$V(\phi)$}
\Text(120,5)[tr]{$\phi$}
\Line(8,50)(10,50)
\Text(5,50)[r]{$M_s^4$}
\Text(60,-20)[t]{a)}
\SetWidth{1.0}
\Curve{(0,20)(10,10)(20,20)(30,50)(45,60)(50,57)(55,54.5)
(60,53)(80,51)(110,50)}
\end{picture}
\begin{picture}(200,200)(0,0) 
\SetOffset(20,10)
\LongArrow(0,10)(120,10)
\LongArrow(10,0)(10,90)
\Text(15,90)[tl]{$\phi(r)$}
\Text(120,5)[tr]{$r$}
\Text(60,-20)[t]{b)}
\Text(5,60)[r]{$\phi_0$}
\Line(95,10)(95,8)
\Text(95,5)[t]{$R$}
\SetWidth{1.0}
\Curve{(10,60)(30,55)(50,42)(70,25)(80,18)(90,12)(100,10.5)(110,10)}
\end{picture}
\end{center}
\caption{a)~The scalar potential leading to the existence of
absolutely stable Q-balls. b)~The Q-ball profile. }
\end{figure}

Among known forms of baryonic matter Q-balls are the most
energetically favorable one. Being absolutely stable, baryonic Q-balls
are viable candidates for the dark matter\cite{KS}. They can be
produced in cosmologically significant amounts in the early
Universe\cite{KS} through the decay of the Affleck-Dine
condensate\cite{AD}. In this talk which is based on
refs.\cite{KKST,KSTT,KST} we discuss modes of detection of relic
Q-balls, as well as their possible role in the evolution of neutron
stars.

\paragraph{2.~Experimental detection of Q-balls.} 
If exist, relic Q-balls are concentrated in galactic halos and
have velocities of order $v\sim 10^{-3}c$. Assuming that Q-balls constitute
cold dark matter with $\rho_{_{DM}} \approx 0.3$~GeV/cm$^3$ one
finds their number density 
\begin{equation}
n_{_Q}\sim \frac{\rho_{_{DM}}}{M_{_Q}} \sim 3 \times 10^{-4} \, Q^{-3/4} 
\left ( \frac{1 {\rm TeV}}{M_s}
\right ) {\rm cm}^{-3}, 
\label{num_dens}
\end{equation}
and corresponding flux 
\[
F\sim n_{_Q} v \sim 3 \times 10^{11} \, Q^{-3/4} 
\left ( \frac{1 {\rm TeV}}{M_s} \right ) {\rm cm}^{-2} {\rm yr}^{-1}. 
\]

Consider the interaction of baryonic Q-balls with ordinary matter. The
interior of a baryonic Q-ball is characterized by large VEV of certain
squark (and, possibly, slepton and Higgs) fields. Therefore, color
SU(3) is broken and de-confinement takes place inside a
Q-ball. Simultaneously, quarks and possibly some leptons get masses
through mixing with gluinos. Since masses are proportional to VEV's
the outer region of a baryonic Q-ball has a layer where quarks are
lighter than $\Lambda_{_{QCD}}$ and are not confined. When a nucleon
enters this region, it dissociates into quarks which later get
absorbed into condensate via the reaction $qq\to\tilde q\tilde q$. The
cross section of this reaction which goes through the gluino exchange
can be parameterized as $\sigma(qq\to\tilde q\tilde q) = \beta/M_s^2$,
where $\beta$ is dimensionless parameter to be treated
phenomenologically.  In total, the reaction looks as
\begin{equation}
(Q) + N \to (Q+1) + \mbox{pions},
\label{Q->Q+1}
\end{equation}
where $N$ denotes a nucleon and it is assumed that, like in typical
hadronic process, the released energy (of order 1 GeV per nucleon) is
carried out predominantly by pions.

Since nuclei in ordinary matter are electrically charged, a Q-ball
resulting from the reaction (\ref{Q->Q+1}) is (with some probability)
positively charged and further absorption is suppressed by the Coulomb
barrier. Depending on their ability to retain electric charge, Q-balls
associated with different flat directions of the MSSM can be divided
into two general classes: Supersymmetric Electrically Neutral Solitons
(SENS) which rapidly neutralize, and Supersymmetric Electrically
Charged Solitons (SECS) which stay charged much longer than the time
between successive collisions. An example of SECS is a Q-ball
associated with $(QQQLLLe)$ flat direction. A non-zero VEV of both
left ($L$) and right ($e$) selectron along this direction makes
electron heavy so that it is repelled by the Q-ball.

{\tolerance=1000 
The cross section of the reaction (\ref{Q->Q+1}) is determined by the
Q-ball size $R$, $\sigma \sim 10^{-33} Q^{1/2} \linebreak (1 {\rm
TeV}/m)^2$~cm$^{2}$. With this cross section, a SENS passing through
ordinary matter with density $\rho$ experiences roughly $100\times
(Q/10^{24})^{1/2} \rho/(1\mbox{~g/cm}^3)$ collisions with nuclei per
centimeter of track. Large amount of energy (of order $100\times
(Q/10^{24})^{1/2}$ GeV/cm) released in pions
is the main signature of these events.

}

Events produced by SECS would look totally different. After few first
collisions SECS becomes electrically charged; from that point on its
interaction with nuclei becomes elastic. The corresponding cross
section is determined by the Bohr radius, $\sigma\sim \pi r_B^2\sim
10^{-16}$cm$^2$.  SECS propagation through matter results in similar
energy release, $\sim 100$ GeV/cm, but now mainly in the form of heat
with only $\sim 10^{-5}$ fraction of visible light.

The present experimental limit on the flux of SECS, $F < 1.1 \times
10^{-14}$ cm$^{-2}$~s$^{-1}$~sr$^{-1}$, is set by the MACRO
search\cite{macro} for ``nuclearites''\cite{dg}, which have similar
interactions with matter.  This translates into the lower limit on the
baryon charge of dark-matter Q-balls, $Q \stackrel{>}{_{\scriptstyle
\sim}} 10^{21}$.  Signatures of SENS are similar to those expected
from the Grand Unified monopoles that catalyze the proton decay.  If
one translates the current experimental limits from
Baikal\cite{baikal} on the monopole flux, one can set a limit on the
charge of SENS, $Q \stackrel{>}{_{\scriptstyle \sim}} 3 \times
10^{22}$, for $m=1~{\rm TeV}$. Of course, this does not preclude the
existence of smaller Q-balls with lower abundances that give
negligible contribution to the matter density of the Universe.

\paragraph{3.~Astrophysical implications.} Since Q-balls are the
most energetically favorable state of baryonic matter, one would
naturally expect them to play role in stellar evolution. This role is,
however, limited by the low density of ordinary matter and small size
of Q-balls. In fact, the only place where they may be essential is the
evolution of neutron stars. 

The neutron star is sufficiently dense to stop both SECS and SENS. We
will not make difference between them in subsequent discussion since
both absorb neutrons with comparable rates. Captured Q-balls reach the
center of the star in a matter of seconds where they finally merge
forming a large central Q-ball. When the charge of the central Q-ball
exceeds $\sim 10^{30}$ further accumulation of relic Q-balls can be
neglected. Therefore, the ultimate fate of the neutron star does not
depend on the flux of Q-balls as long as at least one Q-ball is
captured during the lifetime of a star.

The growth of the central Q-ball in a neutron star is determined by
the infall rate of neutrons on the Q-ball and the rate of processing
of quarks into condensate, whichever is smaller. Since both rates are
proportional to the surface area of Q-ball, the resulting change of the
Q-ball charge obeys the equation
\begin{equation}
dQ/dt = \alpha Q^{1/2}, 
\label{one_q}
\end{equation}
where $\alpha$ is a phenomenological parameter. The magnitude of
$\alpha$ is limited by the infall rate at the level of
$\sim 10^{16}$~s$^{-1}$. With this value of $\alpha$ the neutron star would
live only $\sim 10^5$ years. More realistic value of $\alpha$ is
obtained when the conversion rate is taken into account. Writing the
cross section of the reaction $qq\to \tilde q\tilde q$ as above, one
gets an estimate $\alpha = 10^8 ( M_s/1{\rm TeV})^{-5} \beta \, {\rm
s}^{-1}$, which implies for the lifetime of a star
\begin{equation}
t_s \sim \frac{1}{\beta} \times 
\left ( \frac{m}{200\, {\rm GeV}} \right )^{5} {\rm Gyr}. 
\label{lifetime}
\end{equation}
The lifetime can be as low as $10^{-2}$ Gyr for $m \sim 100$ GeV, or
can exceed the age of the universe for $m \stackrel{>}{_{\scriptstyle
\sim}} 300$ GeV, if $\beta \sim 1$. Since neutron stars with ages
around 0.1 Gyr are known to exist, there is a lower bound on $t_s$
and, correspondingly, on $m^5/\beta$.  If $\beta \ll 1$ and $m$ is in
the TeV range then the lifetime of a neutron star exceeds the age of
the Universe and the relic Q-balls play no role in the stellar
evolution at present time.

Neutron stars are only stable in a certain range of
masses\cite{st}. When the mass of a star becomes smaller than $\sim
0.2 M_{\odot}$ (the mass and gravitational effect of the Q-ball can be
neglected), the star becomes unstable and explodes. The energy
released in this process is of order $\sim 10^{52}...10^{53}$ erg. The
emission of gamma rays associated with the explosion can, in
principle, account for observed gamma-ray bursts. Whether this
mechanism can also explain the duration and spectrum of observed gamma
ray bursts remains to be seen.

\paragraph{Acknowledgments.} I would like to thank the organizers of 
XXIII Rencontres de Moriond. This work is supported in part by INTAS
grant \#INTAS-94-2352 and Russian Foundation for Fundamental Research,
grant \#96-02-17804a.

\end{document}